\documentclass[a4paper,12pt]{article}
\pdfoutput=1
\usepackage{epsfig}
\usepackage{amssymb}
\usepackage{subcaption}
\usepackage{amsfonts}
\usepackage{amsmath}
\usepackage{euscript}
\usepackage{xcolor}
\usepackage{verbatim}
\usepackage{commath}
\usepackage{latexsym}
\usepackage{graphicx}
\usepackage{caption}
\usepackage{float}
\usepackage{bigints}
\usepackage{subcaption}
\usepackage{bbm}
\usepackage{subfiles}
\usepackage{listings}

\usepackage{booktabs}

\newcommand{\beq}{\begin{equation}}

 \def\IR{{\mathbb R}}
 
\def\s{\text{ }}
\def\tb{\Tilde{\beta}}

\newif\ifdtup

\catcode`\@=11

\@addtoreset{equation}{section}

\def\@normalsize{\@setsize\normalsize{15pt}\xiipt\@xiipt
\abovedisplayskip 14pt plus3pt minus3pt%
\belowdisplayskip \abovedisplayskip
\abovedisplayshortskip \z@ plus3pt%
\belowdisplayshortskip 7pt plus3.5pt minus0pt}

\def\small{\@setsize\small{13.6pt}\xipt\@xipt
\abovedisplayskip 13pt plus3pt minus3pt%
\belowdisplayskip \abovedisplayskip
\abovedisplayshortskip \z@ plus3pt%
\belowdisplayshortskip 7pt plus3.5pt minus0pt
\def\@listi{\parsep 4.5pt plus 2pt minus 1pt
     \itemsep \parsep
     \topsep 9pt plus 3pt minus 3pt}}

\relax

\catcode`@=12

\catcode`\@=11

\def\section{\@startsection{section}{1}{\z@}{3.5ex plus 1ex minus
   .2ex}{2.3ex plus .2ex}{\large\bf}}

\newcommand{\bea}{\begin{eqnarray}}
\newcommand{\eea}{\end{eqnarray}}

\def\e{Equation }
\def\r{\right}
\def\l{\left}
\def\s{\text{  }}

\def\SymBoxes#1#2#3#4{\newdimen\un@t \un@t#3%
\raisebox{#1}{\rule{#2\un@t}{#4}\hskip-#2\un@t
\@tempdimb\un@t \advance\@tempdimb by-#4\@tempcntb#2\relax%
\@whilenum{\@tempcntb>0}\do{
\rule{#4}{\un@t}\hskip\@tempdimb \advance\@tempcntb by\m@ne}%
\hskip-#2\un@t \rule[\un@t]{#2\un@t}{#4}%
\rule[\un@t]{#4}{#4}\hskip-#4
\rule{#4}{\un@t}}\hskip-#4}                

\begin{document}


\begin{titlepage}
\begin{flushright}
\end{flushright}
\bigskip
\def\thefootnote{\fnsymbol{footnote}}

\begin{center}
{\Large {\bf A Vestige of FZZ Duality in Higher Dimensions}}
\end{center}

\bigskip
\begin{center}

Chethan Krishnan$^a$\footnote{\texttt{chethan.krishnan@gmail.com}}, \ \&  \ Sarthak Talukdar$^a$\footnote{\texttt{sarthak.talukdar.physics@gmail.com}}
\vspace{0.1in}

\end{center}

\renewcommand{\thefootnote}{\arabic{footnote}}

\begin{center}

$^a$ {Center for High Energy Physics,\\
Indian Institute of Science, Bangalore 560012, India}\\

\end{center}

\noindent

\begin{center} {\bf Abstract} \end{center}
In 1+1 dimensions, the equations of motion of the Horowitz-Polchinski (HP) effective string have a re-writing in terms of a first order system. This is attributed to FZZ duality. In this note, we observe that a similar re-writing exists in higher dimensions, so that the degree of the dilaton-winding subsystem reduces to first order. The 1+1 first order equations emerge as a natural limit of the higher dimensional HP system in the cap region of the cigar. As a result, there is a critical value of the winding amplitude that matches with the 1+1 coset SCFT prediction. At this critical point, the cigar has a puncture at the Euclidean horizon and the {\em higher dimensional} black hole entropy is correctly reproduced by the winding  condensate.

\vspace{1.6 cm}

\vfill

\end{titlepage}

\setcounter{footnote}{0}
\tableofcontents
\section{Introduction}

The connection between Euclidean and Lorentzian black holes has been a source of both insights and puzzles ever since the original work of Gibbons and Hawking \cite{GibHawk}. The Euclidean approach provides a natural path to the derivation of black hole thermodynamics, while obscuring its statistical origins. The Euclidean gravity path integral is not well-defined due to the conformal mode \cite{HawkingPerry}, but its saddles successfully capture the equilibrium physics of quantum gravity.

One may hope that this puzzling situation has improved somewhat in the recent past, thanks to the successful calculation of the unitary black hole Page curve using Euclidean methods\footnote{...and (importantly!) the assumption of local equilibrium during each epoch of Hawking radiation.} \cite{Penington, AEMM, PSSY, MaldaGroup}. It is to be noted however, that the Page curve can be correctly reproduced in general quantum systems irrespective of various details, if the time evolution satisfies a local equilibrium approximation. This fact was demonstrated in detail in \cite{Liu, Vyshnav}. See Appendix A of \cite{Vaibhav2} for a pedagogical presentation that makes a precise parallel between general quantum systems and the calculation in \cite{PSSY}. Since we already knew \cite{GibHawk} that Euclidean gravity can reproduce equilibrium physics, this means that we have to be careful in deciding what exactly we have learned regarding Euclidean gravity from the recent Page curve calculations\footnote{Perhaps the most important lesson we can learn is not {\em about} Euclidean path integrals, but via Euclidean path integrals about the utility and interpretation of quantum extremal surfaces \cite{EngelhardtWall} in Lorentzian signature.}. In particular, the factorization problem is only as much a problem in holography as it is in ordinary systems and their Page curves, where replicas are useful due to approximate equilibrium.

In light of these observations, it may be worth taking a step back and remembering that there is a manifest distinction between Euclidean and Lorentzian black holes in string theory. This follows from the old work of Sathiapalan \cite{Sathiapalan}, Kogan \cite{Kogan} and Atick \& Witten \cite{AtickWitten}, and it has been elaborated by many authors since (see in particular \cite{HP}). The key observation is that the Euclidean black hole geometry has a shrinking time circle around which a closed string can slip off. In terms of the spectrum of the string, this means that the winding modes of a string can become light and may therefore condense. This picture is quite intuitive in Euclidean signature, see e.g., \cite{WittenBH, Spenta, Itzhaki-complementarity, Itzhaki-tip-of-cigar, StringyHorizons} for various results in 1+1 dimensions. But what it implies for Lorentzian signature has never been very clear. While intriguing ideas exist \cite{GiribetNunez1, GiribetNunez2, Itzhaki-interiors, Itzhaki-stringy-information}, it is fair to say that the final implications (for physical, Lorentzian, higher dimensional, ... black holes) are yet to be understood.

There are two reasons for our interest in the Euclidean 1+1 dimensional black hole \cite{RabinoviciForges, Spenta}. Firstly, it is an exact string theory background where the worldsheet CFT can be described using an $SL(2,\IR)_k/ U(1)$ coset WZW model \cite{WittenBH}, with the level $k$ related to the radius of the Euclidean time circle. The precise description of the coset changes somewhat depending on signature and also upon whether we are working with the superstring or the bosonic string. The Euclidean and Lorentzian geometries correspond to different cosets, but we will not emphasize it. We will use results from the Euclidean superconformal coset and its FZZ dual \cite{StringyHorizonsII, ItzhakiHartleHawking, ItzhakiPuncture}. The 1+1 dimensional black hole background does not receive perturbative corrections in $1/k$ (viewed as $\alpha'$ corrections) in the superconformal case.

A second reason for interest in the 1+1 dimensional black hole is that it emerges (together with an $S^3 \times T^5$) in the near-horizon limit of a stack of $k$ near-extremal NS5-branes \cite{MaldacenaStromingerNS5}. It is perhaps not too unreasonable to hope that the lessons one learns from this system may translate to the case of the Schwarzschild black hole in 3+1 dimensions. Some of our results in this paper will show that this hope is at least partially correct.

A key fact about the cigar coset is that it is conjecturally dual to the Sine-Liouville CFT. This is called FZZ duality \cite{KKK, KKKReview}. One consequence of the duality will be important to us. It has been argued  \cite{StringyHorizonsII, ItzhakiPuncture} that the winding mode and the zero mode of the metric-dilaton system (which is responsible for shifting the tip of the cigar), are correlated. This lead the authors of \cite{ItzhakiPuncture} to suggest that the Horowitz-Polchinski (HP) effective string equations of motion for the 1+1 dimensional cigar, simplify. Usually, the HP system \cite{HP} is viewed as a description of the Euclidean black hole that incorporates the winding string mode near the Hagedorn temperature. The authors of \cite{ItzhakiPuncture} instead suggested, based on FZZ duality, that the 1+1 dimensional HP system allows a re-writing in terms of a first order system of ODEs and that this first order system is valid even when the temperature of the black hole is small. Numerically solving these equations, a critical value of the winding condensate was identified from a bulk HP calculation. At precisely the critical value, the numerical bulk result of the winding condensate matched non-trivially with an analytic coset SCFT prediction from the worldsheet theory. Furthermore, the cigar develops a puncture on the tip exactly at the critical point. A final interesting observation was that the entropy of the 1+1 dimensional black hole at this critical value, was precisely its Bekenstein-Hawking entropy. 

These observations are remarkable. They also give us a window to test whether the 1+1 black hole in string theory has lessons that generalize to higher dimensions. The HP system is not limited to 1+1 dimensions, and so we can investigate whether the equations of motion resulting from the higher dimensional HP action allow a re-writing of its dilaton-winding sector in terms of first order ODEs. If this is true, it would be evidence that the physics of FZZ duality has vestiges even in higher dimensions. In this paper, we will show that this is indeed the case. We will furthermore find that the resulting equations have natural connections (in the low temperature limit) to the 1+1 HP system. This allows us to make a precise connection between higher dimensional black holes, and the critical winding condensate (and a puncture at the tip) predicted by the 1+1 coset SCFT. A noteworthy feature of this calculation is that the entropy carried by the winding condensate is that of the {\em higher dimensional} (and not 1+1) black hole. This suggests that the cap-region physics, even in higher dimensions, is effectively captured by the physics underlying FZZ duality. More broadly, these results provide some evidence that the mechanism of winding string condensation may be of significance even in higher dimensions in understanding the Euclidean horizon.

In the next section, we briefly review the 1+1 Horowitz-Polchinski action \cite{ItzhakiPuncture} and the solution space of the first order system that emerges from it. We emphasize two points which go somewhat beyond the discussion in \cite{ItzhakiPuncture}. Firstly, we solve the system at large but finite $k$ and demonstrate that the qualitative asymptotic behavior expected from the analytic solution (where one treats the winding as a probe), emerges from the fully backreacted numerical solutions. This includes the linear dilaton behavior at large distances. (The numerical solutions in \cite{ItzhakiPuncture} were for $k = \infty$ and therefore only applied to the cap region.) We also emphasize that at finite-$k$, the critical value of the winding condensate obtained from the numerical solutions, matches with the string prediction only approximately. 
This mismatch at finite $k$ is an indication that one has to be more careful with finite $\alpha'$-effects in the HP system, if one is interested in ``beyond-the-cap" physics.

In Section \ref{dHPsec} we present the higher dimensional (in general $D >3 $) action and equations of motion of the HP effective string.  We explicitly show that the 2-derivative equations of motion allow a re-writing in which the equations for the winding mode, the dilaton and the temporal component of the metric are first order. A new feature in higher dimensions is the presence of the metric component on the sphere. We will see nonetheless that there is a natural low-temperature limit where the dynamics of the cap simplifies and makes connections with the 1+1 first order system. This enables us to identify the critical solution and condensate and the puncture at the tip. The entropy carried by winding at the critical/punctured cigar is precisely the Schwarzschild black hole entropy, as we show. We conclude with some open questions in a final section, and relegate some technical details to the Appendices.

\section{Horowitz-Polchinski in 1+1 Dimensions}\label{HP2}

In this section, we review and extend the Horowitz-Polchinski system \cite{HP} analyzed in 1+1 dimensions in \cite{ItzhakiPuncture}. We find numerical solutions of the first order HP system at infinite $k$, and also large but finite $k$. These latter solutions start at the horizon and extend all the way to the asymptotic region (the region beyond the cap). By carefully discussing the numerical errors associated to these solutions, we argue that the match between the critical condensate and the string calculation is a large-$k$ phenomenon. At finite $k$, while critical solutions do exist, the condensate value is lower than the string prediction, with the match becoming increasingly tighter as $k$ is increased. We speculate that this mismatch is a result of not keeping track of the higher $\alpha'$-effects in the HP effective string action.

\subsection{The Cap Solution}
\label{itzhakireview}

The Euclidean version of the $SL(2,\mathbb{R})_k/U(1)$ coset CFT describes string theory on the cigar geometry \cite{StringyHorizonsII} with metric and dilaton
\begin{equation} \label{cigar}
    ds^2 = \l(d\rho^2+k \tanh^2{\frac{\rho}{\sqrt{k}}}\s d \theta^2\r) \qquad
    \Phi = \Phi_0 - \ln{\cosh{\frac{\rho}{\sqrt{k}}}}.
\end{equation}
We have set $\alpha'=1$. The angular coordinate $\theta \sim \theta + 2 \pi$ is a rescaled Euclidean time direction, while $\rho$ is the direction along the cigar, $\rho=0$ being the tip. String coupling $e^{\Phi}$ tends to zero as a decreasing linear dilaton far from the tip, and reaches its maximum value $e^{\Phi_0}$ at $\rho=0$. The mass of the black hole is determined by $\Phi_0$, while $k$ is a free parameter that controls the overall size of the cigar. If $k$ is large, this is a weakly curved geometry. As mentioned in the introduction, the worldsheet action with the above background fields (note that the tachyon is zero, if one views this as a bosonic string background) can be written as a gauged WZW model at level $k$ \cite{RabinoviciReview}, and therefore is manifestly a CFT. 

\begin{figure}[h]
     \centering
     \begin{subfigure}[b]{0.4\textwidth}
         \centering
         \includegraphics[width=\textwidth]{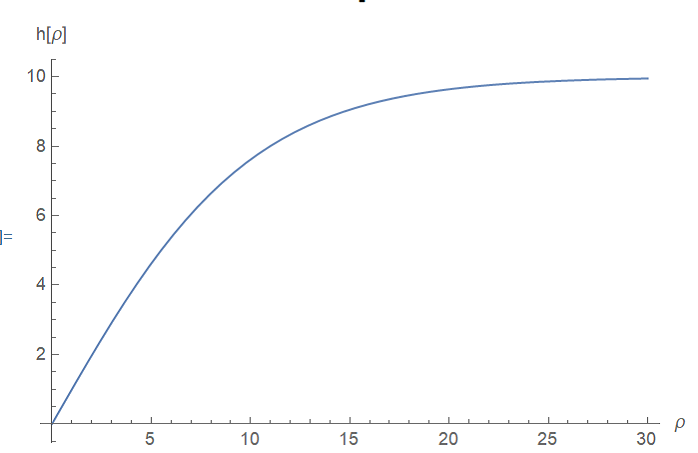}
         \caption{$\sqrt{k} \tanh{\frac{\rho}{\sqrt{k}}}$}
         \label{fig:hexact}
     \end{subfigure}
     \hfill
     \begin{subfigure}[b]{0.4\textwidth}
         \centering
         \includegraphics[width=\textwidth]{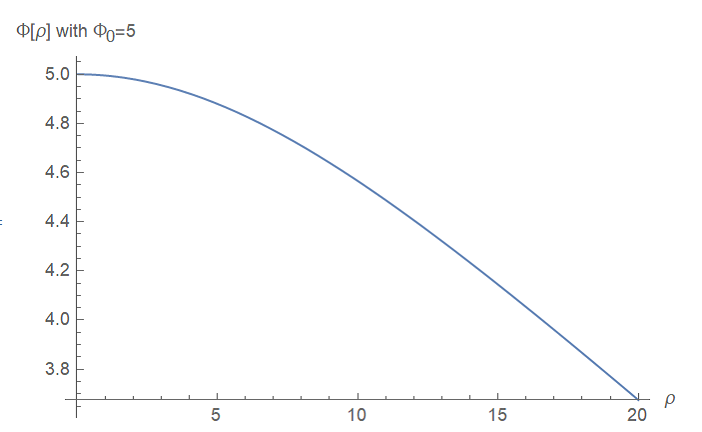}
         \caption{$\Phi_0 - \ln{\cosh{\frac{\rho}{\sqrt{k}}}}$}
         \label{fig:g4n}
     \end{subfigure}
     \caption{Geometry of \e (\ref{cigar}) with $k = 100$. Observe that the cap region of the cigar (i.e., where the metric function is growing linearly and $\Phi$ is non-linear) extends up to $\rho \sim \sqrt{k} = 10$. When $k \rightarrow \infty$, this cap region extends to infinity.}
     \label{fig:cftdescription}
\end{figure}

We will be interested in studying the impact of winding mode physics on this Euclidean black hole. Horowitz and Polchinski, following \cite{Sathiapalan, Kogan, AtickWitten}, wrote down an effective Euclidean theory on the thermal manifold that incorporates the possibility of light winding modes at the Hagedorn temperature. The idea is that once you are in Euclidean signature, nothing prevents us from viewing the path integral/partition function with a Euclidean time circle as a zero temperature partition function on a spatial manifold with a compact circle. The Hagedorn divergence of the partition function can then be attributed to some fields becoming massless in this revised interpretation of the theory when the compactification radius reaches the Hagedorn temperature. A key observation is that because of the anti-periodic (Neveu-Schwarz) boundary conditions for fermions along the thermal circle, the zero energy is tachyonic even in superstring theories. From the spectrum of the (type II) string, we have
\bea
m^2 = \frac{N}{\alpha'} + \frac{4 \pi^2n^2}{\beta^2} + \frac{w^2 \beta^2}{4 \pi^2 \alpha'^2} - \frac{2}{\alpha'} 
\eea
where we use $N$ to denote the collective oscillator contribution, $n$ is the momentum quantization on the circle, and $w$ is the winding quantum number. Since we know that the Hagedorn transition happens at $\beta_H \sim \alpha'^{1/2}$, it follows that the natural candidate for the massless fields are the $w=\pm 1$ winding modes (the higher winding modes become massless at super-Hagedorn temperatures where the partition function does not exist anyway). Writing this pair of modes as a complex field $\chi$ (sometimes called a thermal scalar), it is clear that the partition function contains the piece $Z_\chi = \int D\chi \exp(-S_\chi)$ with the ``effective action" \cite{HP}
\bea
S_\chi= \beta \int d^dx \ (\partial_i \chi \partial _i \chi^* + m^2(\beta) \chi \chi^*)
\eea
where 
\bea
m^2(\beta) = \frac{\beta^2-\beta_H^2}{4 \pi^2 \alpha'^2}, \ \ {\rm with} \ \ \beta_H =  2 \sqrt{2} \pi \sqrt{\alpha'}.
\eea
We have temporarily re-instated the $\alpha'$ for clarity. It is clear that this mode is exactly massless at the Hagedorn temperature. 

The above discussion applies to string theory in flat space. For the curved background in \eqref{cigar} the natural generalization is of the form \cite{HP, ItzhakiPuncture}
\begin{eqnarray}
\label{HP}
I=\int d^2 x \sqrt{g}e^{-2\Phi}\left(-\frac{1}{2\kappa^2}(R-2\Lambda+4\partial^{\mu} \Phi \partial_{\mu} \Phi)+ \partial^{\mu}\chi \partial_{\mu}\chi^* + \frac{\beta^2 g_{\tau \tau}-\beta_H ^2}{4\pi^2 }\chi \chi^*\right),
\end{eqnarray}
where we have incorporated all the (potentially) massless fields. Here $\kappa^2 = 8\pi G_N$, $\Phi$ is the dilaton, $g_{\mu\nu}$ is the metric and $\chi,\chi^*$ are the $\pm 1$ winding modes.
We have again reverted to the unit $\alpha'=1$, with $\beta=2\pi\sqrt{k}, \beta_H = 2\pi \sqrt{2}$, $\Lambda=-\frac{2}{k}$. The cosmological constant is necessary to admit the solution \eqref{cigar}. 

We will consider metrics of the form 
\begin{equation}\label{ansatz}
ds^2=h^2(\rho)d\tau^2+d\rho^2~,
\end{equation}
where $\rho$ is the radial direction, $\tau$ is the thermal circle,
$\tau\sim\tau+\beta$ and the dilaton and winding mode are functions of only the radial coordinate, $\Phi(\rho)$ and $\chi(\rho)$, respectively. After the rescaling
\begin{equation}
\sqrt{k} h \to h \qquad \kappa \chi \to \chi
\end{equation}
the equations of motion arising from \eqref{HP} are \cite{ItzhakiPuncture}
\begin{align}
\label{bad2}
h\left(\frac{\Phi'}{h}\right)'  = & (\chi')^2+h^2\chi^2 \nonumber \\
h\chi ''+h'\chi ' -2 h\chi ' \Phi' =&(h^2 - 2) h \chi \\
\frac{2}{k}+2\Phi''-2(\Phi')^2 =& (\chi')^2 + (3h^2-2)\chi^2 \nonumber
\end{align}
Superficially, our arguments motivating the HP action suggest that it should be reliable only when the temperature $\sim 1/\sqrt{k}$ is close to the Hagedorn temperature $\sim 1$. When the temperature is small (i.e., $k$ large) it is not clear that the HP equations should apply. The authors of \cite{ItzhakiPuncture} argue that the situation may in fact be better, because of FZZ duality. Since the two sides of the duality relate the winding mode to the graviton-dilaton zero mode, they argue that HP effective string description should retain both in order to be consistent. As a result, the higher string modes and the higher derivative corrections to the action were ignored in \cite{ItzhakiPuncture}, while using the system at temperatures far below the Hagedorn temperature. It was noted in \cite{ItzhakiPuncture} that the HP equations allows a re-writing in terms of a first order system at arbitrary $k$, and it was used as an argument for neglecting the higher terms. 

More concretely, the existence of FZZ duality was used as an indication that the dynamics of the various HP modes should not truly be independent, at the level of the equations of motion. It turns out that this is indeed the case. It was shown in \cite{ItzhakiPuncture} that there exists a first order system of equations which can ``solve" the second order equations that we have written above:
\begin{align}
\label{good2}
h' &= h \Phi' +1  \nonumber \\
\Phi' &= -h\left(\chi^2+\frac{1}{k} \right)  \\
\chi ' &= -h \chi \nonumber
\end{align}
We will call this the first order re-writing of \eqref{bad2}. The claim then, is effectively that the FZZ-compatible dynamics of the HP equations is captured by this first order system.

In \cite{ItzhakiPuncture} solutions of this system in the $k\rightarrow \infty$ limit were obtained. The idea is to treat the winding mode as a probe at large enough $\rho$ to obtain reasonable boundary conditions, and numerically integrate towards the horizon using the full first order HP system. It is easy to check that when $\chi=0$, the system reproduces the 1+1 d black hole as the background solution:
\begin{eqnarray}\label{BC2}
h(\rho)=\sqrt{k}\tanh\left(\frac{\rho}{\sqrt{k}}\right)
\qquad
e^{\Phi}=\frac{e^{\Phi_0}}{\cosh\frac{\rho}{\sqrt{k}}}
\end{eqnarray}
Using \eqref{BC2} in the first order system \eqref{good2} and solving for $\chi$ yields
\begin{equation}\label{coshk}
\chi(\rho)=\frac{A}{\cosh^k\left(\frac{\rho}{\sqrt{k}}\right)}
\end{equation}
This is the probe solution, which can be used to set boundary conditions for numerical integration for the full backreacted system. 

When $k$ is sufficiently large, we can set boundary conditions at $1 \ll \rho = \Tilde{\rho}\ll \sqrt{k}$. Note that because $k$ is taken to be sufficiently large, we can work with large $\rho$ (in relation to string length which is unity) while still being inside the cap. In this regime, \eqref{BC2} and \eqref{coshk} can be approximated by
\begin{equation}
\label{chGauss}
\chi(\rho=\Tilde{\rho})=A \s e^{-\frac{\Tilde{\rho}^2}{2}}, \qquad
h(\rho=\Tilde{\rho})=\Tilde{\rho}, \qquad \Phi'(\rho=\Tilde{\rho})=0.
\end{equation}
This is 2-dimensional Euclidean flat space with a constant dilaton, and a Gaussian decaying winding mode. Note that the scale of the Gaussian exponent is set by $\alpha'$ which is our unit. We have found that picking a location to start the integration from, is somewhat delicate. In order to get the successful results of \cite{ItzhakiPuncture} we have to start at $\tilde \rho \sim 5$ which is a reasonable trade-off between numerical errors in Mathematica and  (expectations for) backreaction errors. Our Mathematica errors in this case are of $\mathcal{O}(10^{-11})$, and from \eqref{HP} we expect the backreaction errors to be the of the order of $\chi^2 \sim A^2 e^{-\rho^2}$. For larger $\rho$ the backreaction errors are smaller, but the numerical errors are bigger than the backreaction errors and therefore the numerical solution is unable to distinguish $A$ as sharply as it can at $\tilde \rho=5$.

\begin{figure}[h]
    \centering
    \includegraphics[width = 10cm, height = 5cm]{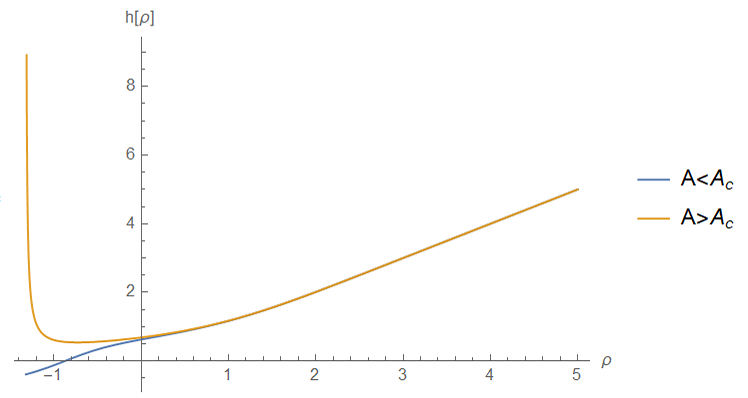}
    \caption{Both curves are for $h(\rho)$, the blue one for $A = 0.74$  below $A_c$, and the orange one for $A = 0.75$ above $A_c$.}
    \label{fig:h}
\end{figure}

\begin{figure}[h]
    \centering
    \includegraphics[width=13cm, height=6cm]{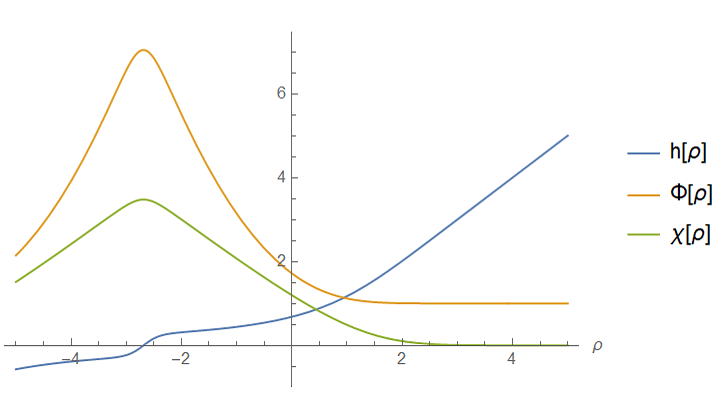}
    \caption{Solutions of $h(\rho), \Phi(\rho)$
 and $\chi(\rho)$ near the horizon with the slightly sub-critical $A = 0.749306$.}
    \label{fig:2dorigin}
\end{figure}

Depending on the value of $A$, there are two classes of solutions. In one class, $h(\rho)$ vanishes at some $\rho_0$. This can be viewed as the tip of the geometry. As the amplitude $A$ approaches a critical amplitude $A_c$ from below, $\rho_0$ becomes more and more negative, and a narrow neck develops. Above the critical $A_c$ there is a second class of solutions, characterized by a vanishing $h'$ at some $\rho_0 <0$ with $h(\rho_0)>0$, as well as $h(\rho$) diverging at a finite negative $\rho < \rho_0$. We can view this as  a puncture at the tip.  Representative plots of both classes are shown in Figure \ref{fig:h}, while a nearly critical sub-critical solution is shown in Figure \ref{fig:2dorigin}.
 \par

In \cite{ItzhakiPuncture}, it was shown that the critical value of $A$ for any $k$ can be obtained from a worldsheet calculation in the SCFT. The result was
\begin{equation}\label{Ac}
    A_c(k) = \left(\frac{\Gamma(1 + \frac{1}{k})}{\Gamma(1 - \frac{1}{k})}\right)^{k/4}
    \qquad
    \Rightarrow
    \qquad
    \lim_{k \to \infty} A_c = e^{-\gamma/2} = 0.749306001\dots
\end{equation}
In the $k \rightarrow \infty$ limit, this matches the critical value of $A$ obtained from the HP solution on the nose. While we will not directly use it here, it is worth noting that the analytic expression for $A_c(k)$ has a perturbative expansion in $1/k$ around the $k=\infty$ value.

\subsection{Beyond the Cap: Finite-$k$ Solutions}
\label{full2}

We find it noteworthy that such a simplification, motivated by FZZ duality, exists. As was shown in \cite{ItzhakiPuncture} and we have partially reviewed, this simplified system has the striking feature that it has a critical solution which precisely matches with the worldsheet CFT prediction. It also reproduces the correct 1+1 black hole entropy from the winding condensate. These in our view are strong suggestions that the relevant metric-dilaton-winding system is indeed aware of FZZ duality. We are also sympathetic to the possibility suggested in \cite{ItzhakiPuncture} that the other modes of the string including the higher winding modes and higher derivative terms may not play an important role in the HP dynamics, at least when $k \rightarrow \infty$. 

While we find it plausible that the metric-dilaton-winding subsystem undergoes a simplification at any $k$, it is not clear to us that all the details of the dynamics are unaffected by the higher modes and higher derivative corrections when $k$ is finite. The reason is that $1/k$ is ultimately a measure of $\alpha'$ effects and curvature corrections. In fact the calculation of \cite{ItzhakiPuncture} is done in the strict $k \rightarrow \infty$ limit. We will show that when $k$ is finite but large, we can get numerically reliable solutions of the 1+1 dimensional HP system \eqref{good2} whose critical value of the condensate only approximately matches the string theory prediction. The discrepancy becomes larger, as $k$ decreases. This suggests that the HP system with the first order re-writing can capture the physics of the cap region (i.e., $k \rightarrow \infty$) but to fully describe the asymptotic region of the black hole, we may need to retain more aspects of the full dynamics. We will give more details of the 1+1 dimensional finite-$k$ calculation below. In later sections, we will also show that an analogous picture emerges in higher dimensions as well. We will be able to obtain the physics of the cap region (in particular, the entropy) from a winding condensate in the higher dimensional HP system as well.

\begin{figure}[h]
     \centering
     \begin{subfigure}[b]{0.4\textwidth}
         \centering
         \includegraphics[width=\textwidth]{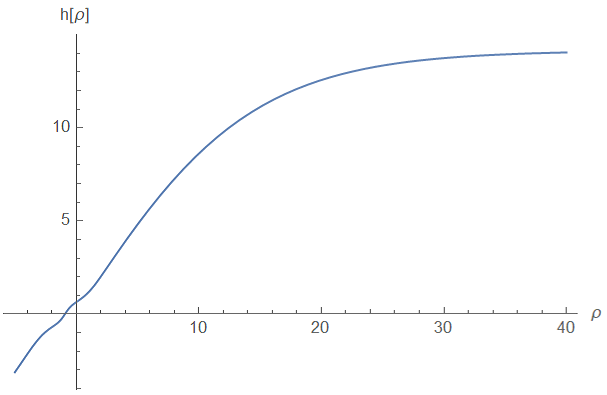}
         \caption{Solution for $h(\rho)$}
         \label{fig:hfar}
     \end{subfigure}
     \hfill
     \begin{subfigure}[b]{0.4\textwidth}
         \centering
         \includegraphics[width=\textwidth]{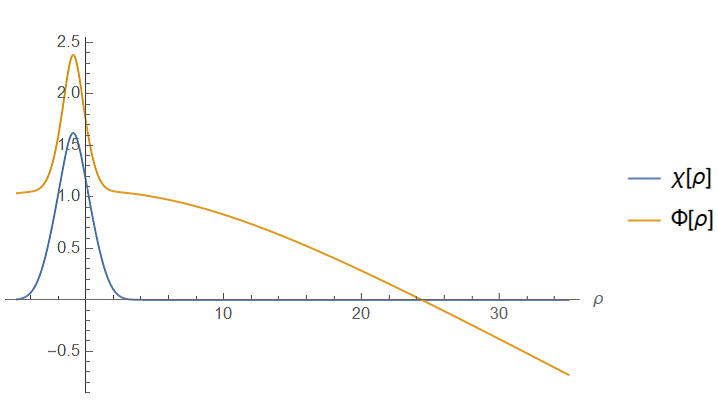}
         \caption{Solution for $\Phi(\rho)$ and $\chi(\rho)$}
         \label{fig:stringfar}
     \end{subfigure}
     \caption{Solution of 1+1 d HP system over the range of $\rho$ starting from horizon to asymptotic region with $k=200$, $\Tilde{\rho}=5$, and $A=0.74$}
     \label{fig:2dfull}
\end{figure}

Getting the full solution is conceptually straightforward, we just have to numerically integrate \eqref{good2} at \emph{finite} $k$. Even thought $k$ is not strictly infinity, we will again start our integrations from around $\tilde \rho =5$, and integrate both inward and outward. The reason we do not start our integrations from farther out, is because the $\chi$ solution is essentially indistinguishable from zero at large $\rho$ and this leads to numerical challenges to putting initial data there. We will again use the probe solution\footnote{Note that this means that we are setting the boundary conditions at scales well-separated from the string scale. This approach will not work, when $k$ is $\mathcal{O}(1)$. This is the usual regime of the HP effective string.} \eqref{BC2} and \eqref{coshk}, to set the boundary conditions at $\rho =\tilde \rho$:
\begin{equation}
\label{bcfull2}
    \chi(\Tilde{\rho}) = \frac{A}{\cosh^k {\frac{\Tilde{\rho}}{\sqrt{k}}}}
    \qquad
    h(\Tilde{\rho}) = \sqrt{k} \tanh{\frac{\Tilde{\rho}}{\sqrt{k}}}
    \qquad
    \Phi'(\Tilde{\rho}) = - \frac{\tanh{\frac{\Tilde{\rho}}{\sqrt{k}}}}{\sqrt{k}}
\end{equation}
and get Figure \ref{fig:2dfull} as the full solution for the 1+1 dimensional HP system, starting from the horizon at $\rho = 0$ to asymptotic region with large $\rho$. It is important to note that this solution, has the features we would like. In the asymptotic region, $h$ saturates and $\Phi$ is a linear dilaton with a negative slope, whereas it behaves like the near-horizon solution we discussed in the previous subsection \cite{ItzhakiPuncture} near the horizon. The linear dilaton behavior in the asymptotic region is a signature of the $SL(2,\mathbb{R})_k/U(1)$ coset CFT.  

\begin{figure}[h]
     \centering
     \begin{subfigure}[b]{0.4\textwidth}
         \centering
         \includegraphics[width=\textwidth]{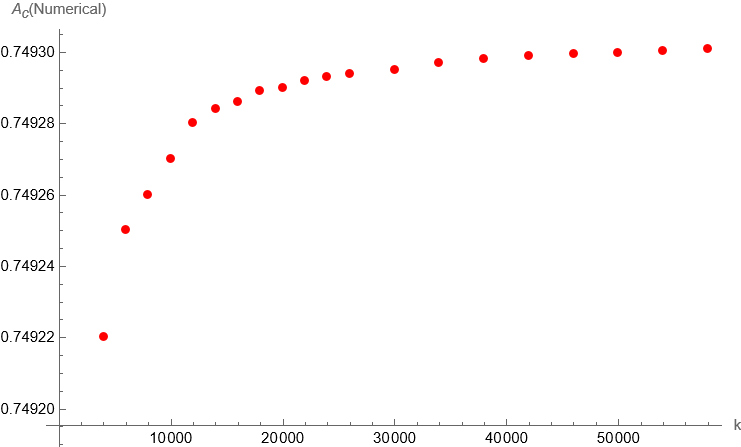}
         \caption{$A_c$ from HP system}
         \label{fig:Acnumeric}
     \end{subfigure}
     \hfill
     \begin{subfigure}[b]{0.4\textwidth}
         \centering
         \includegraphics[width=\textwidth]{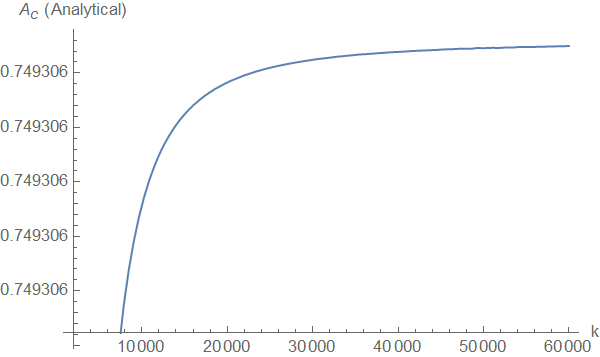}
         \caption{$A_c$ from SCFT description}
         \label{fig:Acanalytic}
     \end{subfigure}
     \caption{$A_c$ values obtained from the numerics of the HP system and the SCFT for various $k$. The qualitative characteristics of both curves match, but it should be noted that the values are systematically smaller in the HP plot. This is more clear in Figure \ref{fig:Acboth}. The numerical values in the HP plots should not be taken too seriously due to precision issues in the final decimal in some cases.}
     \label{fig:Ac}
\end{figure}

\begin{figure}[h]
    \centering
    \includegraphics[width=0.5\linewidth]{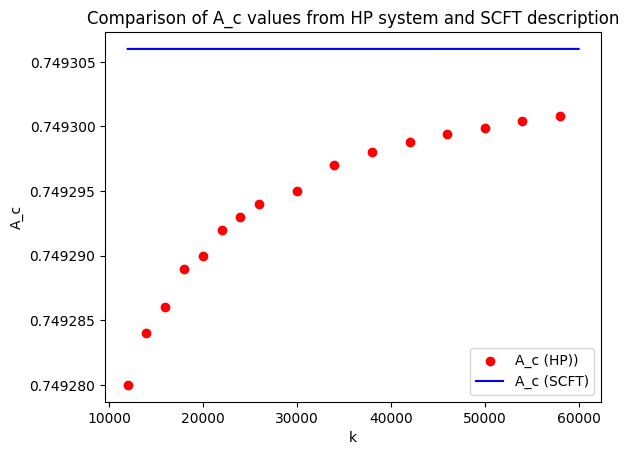}
    \caption{Comparison of $A_c$ values obtained from HP and SCFT. The blue curve looks essentially like a constant, because the decrease in the red curve for the same range of $k$ is hierarchically larger. The numerical values in the red scatter plot should not be taken too seriously, the qualitative features are our emphasis here.}
    \label{fig:Acboth}
\end{figure}

The numerical curves still exhibit the critical behavior, but the critical value of the winding condensate decreases with $k$. This is qualitatively consistent with the SCFT result quoted earlier, but the precise numerical match emerges only as $k\rightarrow \infty$. This is shown in Figure \ref{fig:Acnumeric}. Figure \ref{fig:Acanalytic} is the analytic behavior of $A_c$ as a function of $k$ in \eqref{Ac}. Even though the qualitative nature of the curve matches pretty well, the values start deviating at low values of $k$. In fact the HP result is systematically smaller than the SCFT result at the same $k$ for finite $k$, see Figure \ref{fig:Acboth}.

\begin{figure}[h]
     \centering
     \begin{subfigure}[b]{0.3\textwidth}
         \centering
         \includegraphics[width=\textwidth]{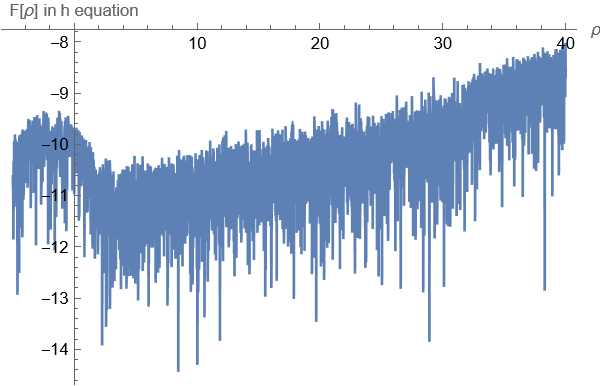}
         \caption{$F(\rho)$ for $h$-equation}
         \label{fig:herr}
     \end{subfigure}
     \hfill
     \begin{subfigure}[b]{0.3\textwidth}
         \centering
         \includegraphics[width=\textwidth]{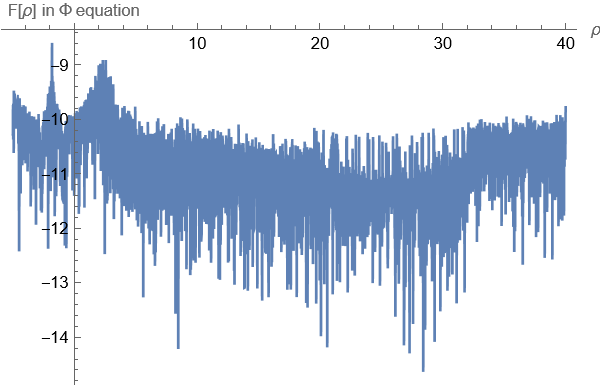}
         \caption{$F(\rho)$ for $\Phi$-equation}
         \label{fig:phierr}
     \end{subfigure}
     \hfill
     \begin{subfigure}[b]{0.3\textwidth}
         \centering
         \includegraphics[width=\textwidth]{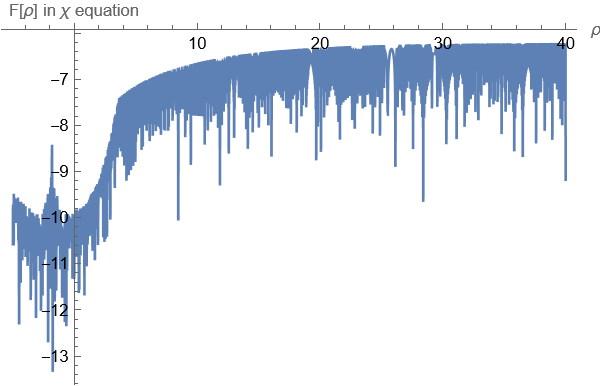}
         \caption{$F(\rho)$ for $\chi$-equation}
         \label{fig:chierr}
     \end{subfigure}
     \caption{Error plots for the $k=200$ case.    }
     \label{fig:error}
\end{figure}

As we mentioned previously, one has to be careful with errors in the numerical integration. So we conclude this section by demonstrating that the numerical errors are well under control in one of the cases\footnote{We have not done an exhaustive error analysis of all the cases (the numerical errors in Mathematica  depend in a non-trivial way on the choice of $k$), but the statement is quite robust and can be checked for many cases.}. We will present the details for the case $k=200$. The numerical HP value for the critical condensate in this case is $A_c^{HP}=0.7477$, while the result from the SCFT is $A_c^{SCFT}=0.7493$. The mismatch is therefore in the 3rd decimal. But the errors in the numerical solutions are in the 6th decimal or later, as can be checked by comparing the left and right hand sides of the HP equations, after computing them numerically from our (numerical) solutions.  We define
\bea
F(\rho) = \log_{10}{\abs{\frac{{\text{LHS} - \text{RHS}}}{\text{LHS}}}}
\eea
and plot this quantity for all three of the equations in \eqref{good2} with $k = 200$ in Figure \ref{fig:error}.
It is clear that the errors are less than $\mathcal{O}(10^{-6})$ at every value of $\rho$ for each of the functions. This shows that the HP plots are reliable for the comparison we are making.

These observations suggests that at finite $k$, while the duality between the winding and the metric-dilaton modes allow a simplification of the HP system, the detailed dynamics may require inclusion of other finite-$\alpha'$ effects. 
Since large $k$ corresponds to large $\beta$, we can expect that understanding the cap region may be possible, even in higher dimensions, using the HP system. We will demonstrate this in the following sections, by first showing that a first order re-writing exists for HP systems even in higher dimensions. 

\section{Horowitz-Polchinski in Higher Dimensions}
\label{dHPsec}

In this section, we first write down the action and equations of motion associated to the $D$-dimensional Horowitz-Polchinski system. It will be shown that these second order equations allow a first order reduction in the relevant sector, analogous to the 1+1 d case. We will view this as a hint of FZZ duality in higher dimensions. We will present the solution for the near-horizon HP solution with $\Lambda = 0$ in 4-dimensions. This is the analogue of the cap region solution in 1+1 dimensions. We will be able to build a precise relation with the 1+1 d equations and identify the critical solution and a puncture at the tip of the cigar.  We will show that the higher dimensional black hole entropy is the winding entropy of this critical solution.

\subsection{Vestige of FZZ duality in $D$ dimensions}
\label{dfzz}

We will be working with the $D$-dimensional metric of the form 
\begin{equation}\label{dmetric}
    G_{MN}^D \s dx^M dx^N =h^2(\rho) \s d\tau^2+d\rho^2+g(\rho)\s d\Omega_{D-2}^2
\end{equation}
where $M$ and $N$ take values from 0 to $D-1$, with the $0^{\rm th}$ dimension being the compact time dimension $\tau$. The Horowitz-Polchinski action in $D$ spacetime dimensions can be written in the form
\bea\label{HPhigh}
    &I = \bigintsss d^Dx \s \sqrt{G_D} e^{-2 \Phi}\biggl(-\frac{1}{2 \kappa_
0^2}\left(R_D-2 \Lambda +4G_D^{MN}\nabla_M \Phi \nabla_N \Phi\right)  + \hspace{1.5in} \nonumber  \\   &\hspace{2in}+G_D^{MN}\nabla_M \chi \nabla_N \chi^*+\left(\frac{\beta^2 G_{\tau \tau}^D-\beta_H^2}{4\pi^2\alpha'^2}\right)\chi \chi^* \biggr) &
\eea
The equations of motion are 
\begin{equation}
\begin{split}
    R-2\Lambda-4\nabla^{\mu}\Phi \nabla_{\mu}\Phi +4 \nabla^2\Phi =   2 \kappa_0^2 \left(\nabla^{\mu}\chi\nabla_{\mu}\chi^*+\frac{\beta^2 G_{\tau \tau}-\beta_H^2}{4\pi^2\alpha'^2}\chi \chi^*\right),
\end{split}
\end{equation}
\begin{equation}
    R_{\tau \tau}+2 \nabla_{\tau} \nabla_{\tau}\Phi=-2\kappa_0^2 \frac{\beta^2 G_{\tau \tau}^2}{4\pi^2\alpha'^2}\chi \chi^*,
\end{equation}
\begin{equation}
    R_{\mu \nu}+2\nabla_{\mu}\nabla_{\nu}\Phi =2 \kappa_0^2 \nabla_{\mu}\chi \nabla_{\nu}\chi^*,
\end{equation} 
\begin{equation}
    \frac{e^{2\Phi}}{\sqrt{G_D}}\nabla_{\mu}\left(\sqrt{G_D}e^{-2\Phi}G^{\mu \nu}\partial_{\nu}\chi \right)=\frac{\beta^2 G_{\tau \tau}-\beta_H^2}{4\pi^2\alpha'^2}\chi.
\end{equation}
Setting the ansatz
\begin{equation}
    \Phi \equiv \Phi(\rho) \qquad \chi \equiv \chi(\rho)
\end{equation}
these become ($\Tilde{\beta} \equiv \frac{\beta}{2 \pi}$)
\begin{equation}\label{deom1}
\begin{split}
    2(\Phi''-\Phi'^2) + \frac{(D-2)(D-3)}{8}\l(\frac{g'}{g}\r)^2 &- \frac{(D-2)(D-3)}{2g}+
    \\
    &- \Lambda+\l(\frac{D-2}{2}\r)\l(\frac{g'h'}{gh}\r) = \chi'^2+\chi^2(3\Tilde{\beta}^2 h^2-\Tilde{\beta_H}^2),
\end{split}   
\end{equation}
\begin{equation}\label{deom2}
    \frac{D-3}{g}- \frac{1}{2}\frac{g'h'}{gh}-\frac{g''}{2g}+\frac{g'\Phi'}{g} - \frac{D-4}{4} \l(\frac{g'}{g}\r)^2=0,
\end{equation}
\begin{equation}\label{deom3}
     \frac{e^{2\Phi}}{g^{\frac{D-2}{2}}h}\partial_{\rho}(g^{\frac{D-2}{2}}h\chi'e^{-2\Phi})=\left(\Tilde{\beta}^2 h^2-\Tilde{\beta_H}^2\right)\chi,
\end{equation}
\begin{equation}\label{deom4}
\begin{split}
    h\left(\frac{\Phi'}{h}\right)'+ \frac{(D-2)(D-3)}{8}\l(\frac{g'}{g}\r)^2 &+ \l(\frac{D-2}{2}\r)\l(\frac{g'h'}{gh}\r) + \\
    &- \frac{(D-2)(D-3)}{2g} - \frac{D-2}{2} \frac{g' \Phi'}{g}=\chi'^2+\Tilde{\beta}^2 h^2\chi^2
\end{split}
\end{equation}
The claim is that the equations for the dilaton-winding-$h$ subsystem can the re-written in a first order form. The manipulations are somewhat more involved due to the presence of the extra field $g(\rho)$ arising from the compact sphere, so we place some comments in Appendix \ref{find1st}. The first-order system in $D$ dimensions after the rescaling $\Tilde{\beta} h \rightarrow h$ is 
\begin{equation}\label{dfinal1}
    h'=h\left(\Phi'- \frac{D-2}{4}\frac{g'}{g}\right)+\frac{\Tilde{\beta_H}^2}{2},
\end{equation}
\begin{equation}\label{dfinal2}
    \chi'=- h\chi,
\end{equation}
\begin{equation}\label{dfinal3}
    \frac{\Phi'}{h}= - \chi^2 + \frac{1}{\Tilde{\beta_H}^2} \left(-\frac{D-2}{8}\left(\frac{g'}{g}\right)^2 +\frac{D-2}{4}\frac{\Tilde{\beta_H}^2}{h}\frac{g'}{g}-\frac{(D-2)(D-3)}{2g} + \Lambda\right).
\end{equation}
To complete the system, we should also include  \eqref{deom2} for $g(\rho)$. The system still has the property that the winding-dilaton-$h$ equations are first order. Note also that the winding does not couple directly to $g(\rho)$ in \eqref{deom2}. We will call these equations the first order system  -- even though the $g$-equation is second order -- because the fields relevant to the FZZ duality are first order. Note in particular that this re-writing is possible before we take any near-cap approximation.

\subsection{The Cap Solution in 3+1 Dimensions}
\label{4nearhorizon}

The first order equations of motion (with $\Lambda = 0$) in 4 dimensions are 
\begin{equation}\label{4final1}
    h'=h\left(\Phi'- \frac{1}{2}\frac{g'}{g}\right)+\frac{\Tilde{\beta_H}^2}{2}
\end{equation}
\begin{equation}\label{4final2}
    \chi'=- h\chi
\end{equation}
\begin{equation}\label{4final3}
    \frac{\Phi'}{h}= - \chi^2 + \frac{1}{\Tilde{\beta_H}^2} \left(-\frac{1}{4}\left(\frac{g'}{g}\right)^2 +\frac{1}{2}\frac{\Tilde{\beta_H}^2}{h}\frac{g'}{g}-\frac{1}{g}\right)
\end{equation}
\begin{equation}\label{4final4}
    \frac{1}{g}- \frac{1}{2}\frac{g'h'}{gh}-\frac{g''}{2g}+\frac{g'\Phi'}{g}=0
\end{equation}

Based on our previous discussion, we expect that we may be able to find a reliable solution for this system in the cap region. This means that we should choose the boundary conditions from the near-horizon region for the Schwarzschild metric. In the standard Schwarzschild metric, 
\begin{equation}\label{schex}
    h(\rho) = \l( 1 - \frac{2m}{r(\rho)}\r)^{1/2} \qquad
    g(\rho) = r^2(\rho) 
\end{equation}
where the usual Schwarzschild coordinate $r$ and our $\rho$ coordinate are related via
\begin{equation}\label{r}
    \rho = \int_{2m}^r \l( 1 - \frac{2m}{r'}\r)^{-1/2} dr' = \sqrt{r(r-2m)}+ 2m \sinh^{-1}{\sqrt{\frac{r}{2m}-1}}
\end{equation}
To avoid the conical singularity at the tip, the periodicity of the imaginary time direction is fixed by the mass of the black hole to be $\beta = 8 \pi m$. When $r \gtrsim 2m$, $r \approx 2m + \frac{\rho^2}{8 m}$. In this near-horizon limit, we find 
\begin{equation}\label{hgexp}
    h(\rho) = \frac{\rho}{\tb} - \frac{\rho^3}{2 \tb^3} + \cdots
    \qquad
    g(\rho) = \frac{\tb^2}{4} + \frac{\rho^2}{2} + \cdots
\end{equation}
Using \eqref{hgexp} in \eqref{4final1} and \eqref{4final4} gives a boundary condition for $\Phi$ in the large $\tilde \beta$ limit of our interest:
\begin{equation}\label{phibc}
    \Phi'(\rho) = 0
\end{equation}
We can now solve for the winding mode by solving \eqref{4final2}. The result is    
\begin{equation}\label{chibc}
    \chi(\rho) = A \s e^{-\frac{\rho^2}{2}}.
\end{equation}
Here, we have found a unique solution from the first-order system. There is also a second-order path to getting this solution. Plugging the leading order terms from \eqref{hgexp} along with \eqref{phibc} into \eqref{deom3} with $D=4$ gives an equation that can be solved for $\chi$. The equation is second order, but by retaining only the normalizable mode, we get the same equation above. A related approach to determining the near-horizon winding mode was taken in \cite{Mertens}.

\begin{figure}[h]
     \centering
     \begin{subfigure}[b]{0.4\textwidth}
         \centering
         \includegraphics[width=\textwidth]{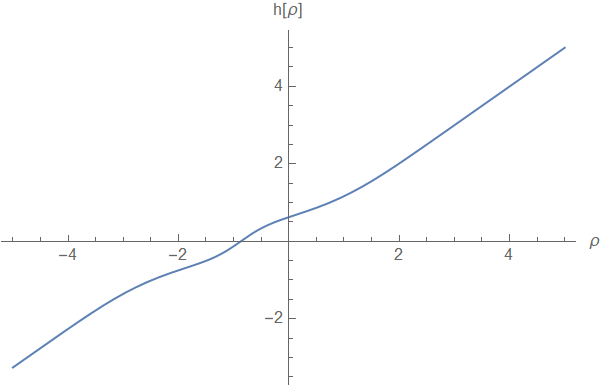}
         \caption{$h(\rho)$}
         \label{fig:h4}
     \end{subfigure}
     \hfill
     \begin{subfigure}[b]{0.4\textwidth}
         \centering
         \includegraphics[width=\textwidth]{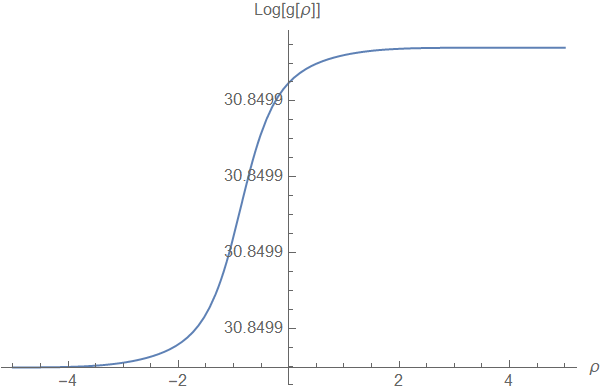}
         \caption{$\log{g(\rho)}$}
         \label{fig:g4}
     \end{subfigure}
     \caption{Near-horizon solution of the 4-dimensional HP system for the metric with $\Tilde{\rho} =5$, $A=0.7$, and $\tb = 10^7$. Note that $\log g$ is roughly constant, the vertical axis is chosen to emphasize the relatively small variation.}
\end{figure}

\begin{figure}[h]
     \centering
     \begin{subfigure}[b]{0.4\textwidth}
         \centering
         \includegraphics[width=\textwidth]{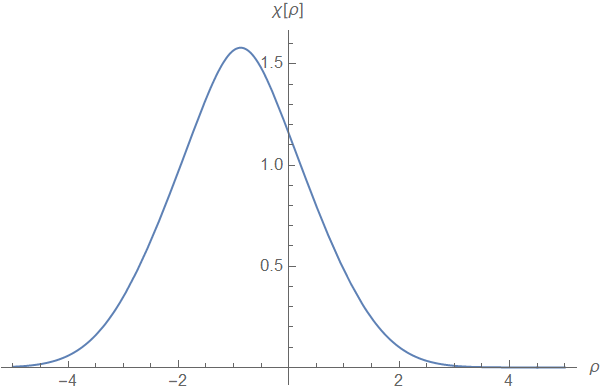}
         \caption{$\chi(\rho)$}
         \label{fig:chi4}
     \end{subfigure}
     \hfill
     \begin{subfigure}[b]{0.4\textwidth}
         \centering
         \includegraphics[width=\textwidth]{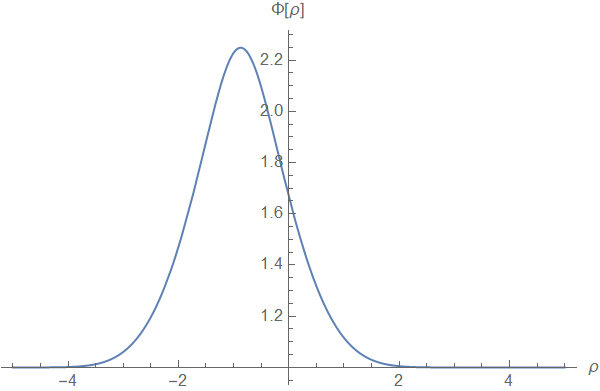}
         \caption{$\Phi(\rho)$}
         \label{fig:phi4}
     \end{subfigure}
     \caption{Near-horizon solution of the 4-dimensional HP system for the winding tachyon and dilaton with $\Tilde{\rho} =5$, $A=0.74$, and $\tb = 10^7$.}
\end{figure}

These calculations allow us to set the boundary conditions for our first-order system in the cap region (note that $\tilde \beta = \beta/2 \pi$): 
\begin{equation}\label{4dbc}
    h(\Tilde{\rho}) = \Tilde{\rho}, \qquad
    g(\Tilde{\rho}) = \frac{\tb^2}{4}, \qquad
    \Phi'(\Tilde{\rho}) = 0, \qquad
    \chi(\Tilde{\rho}) = A \s e^{-\frac{\Tilde{\rho}^2}{2}}.
\end{equation}
We have re-scaled $\tilde \beta h \rightarrow h$ to be consistent with our previous conventions. The analogue of the cap region in 1+1 dimensions here, is the large-$\beta$ regime.  
We can evolve \eqref{4final1}-\eqref{4final4} with the boundary conditions \eqref{4dbc}. The solution is plotted in Figures \ref{fig:h4}, \ref{fig:g4}, \ref{fig:chi4} and \ref{fig:phi4}. 

A crucial observation is that $g(\rho)$ is roughly constant, and when it is large, equations \eqref{4final1}-\eqref{4final4} reduce to the 1+1 first order form. This is a very useful simplification. We find that the cap region solutions for $h$, $\Phi$ and $\chi$ in 3+1 dimensions can be understood via the cap solutions in 1+1 dimensions because of the relative constancy of $g$ at large $\beta$ (which leads to large $g$).  It can also be seen that just like the 2-dimensional case, there is a critical $A_c$ corresponding to a given value of $\tb$ (which is the analogue of given value of $k$). With the increase in $\tb$, $A_c$ increases and approaches precisely the critical value for $A_c$ that was found in \cite{ItzhakiPuncture} in the $k=\infty$ case. This means that there is a direct connection between the large-$\tb$ limit of the 4-dimensional case and the results in \cite{ItzhakiPuncture}. This is intuitive -- we expect that in the large-$\beta$ limit, the higher dimensional geometry reduces to the 1+1 dimensional cap region times a flat geometry because the curvature of the compact sphere becomes negligible. What is interesting here is that this is realized in the emergence and structure of the first order system, thereby giving us a precise way in which the 1+1 dimensional FZZ duality can make statements about the higher dimensional case. 
\begin{figure}[h]
    \centering
    \includegraphics[width=12cm]{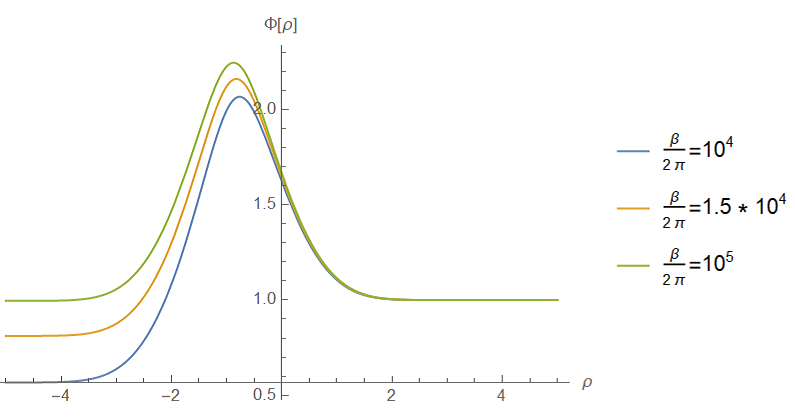}
    \caption{The dilaton solution of the 4-dimensional HP system for different values of $\tb$. The left hand side of the plot shifts upwards and reaches the same level as the right hand side with increase in $\tb$. The latter is a feature of the 1+1 dimensional solution.}
    \label{fig:dilaton}
\end{figure}


\subsection{Winding Entropy as Black Hole Entropy in 3+1 Dimensions}
\label{entropy4d}

The entropy carried by the winding modes can be calculated using 
\begin{equation}
    S_{W} = (\beta \partial_{\beta}-1)I
\end{equation}
where $I$ is given by \eqref{HPhigh} when there is a puncture at the cigar's tip. 

Direct calculation of the action for our ansatz for the fields yields, 
\begin{equation}
    S_W = \frac{2 \beta^3}{(2 \pi \alpha')^2} \bigintsss d^{D-1}x \sqrt{g} e^{-2 \Phi} g_{\tau \tau}| \chi |^2 
\end{equation}
In the $D=4$ case, it becomes
\begin{equation}
\label{swoff}
    S_W = \frac{8 \pi \beta^3}{(2 \pi \alpha')^2} \bigintsss d\rho \s e^{-2 \Phi} h^3 g \chi^2
\end{equation}
We impose the $G_{\tau \tau}$ equation of motion 
\eqref{HPhigh}
\begin{equation}
\label{gtt}
    h'' + \frac{g'h'}{g} - 2 h' \Phi' = \frac{2 \kappa_0^2 \beta^2}{(2 \pi \alpha')^2} h^3 \chi^2
\end{equation}
in \eqref{swoff} to bring things on-shell
\begin{equation}
    \label{swon}
    S_W = \frac{4 \pi \beta}{\kappa_0^2} \bigintsss d\rho \s (h' g e^{-2 \Phi})'
\end{equation}
This is a key step, and the correct power of $\beta$ here is crucial for us to get the correct higher-dimensional black hole entropy. But note that so far we have not used anything specific to the first order system, and these steps were also taken in \cite{Brustein, ItzhakiPuncture}\footnote{Let us emphasize however that since the first order system implies the second order system, \eqref{swon} is a consequence of \eqref{swoff} even if we were to exclusively work with the 1st order system.}. In the former, the entropy was obtained directly in the second order system and in the latter, the subsequent calculations were in the first order system in 1+1 dimensions. We will follow the latter, but in the context of the 3+1 dimensional first order system. We will contrast these calculations in the concluding section.

\begin{figure}[h]
     \centering
     \begin{subfigure}[b]{0.45\textwidth}
         \centering
         \includegraphics[width=\textwidth]{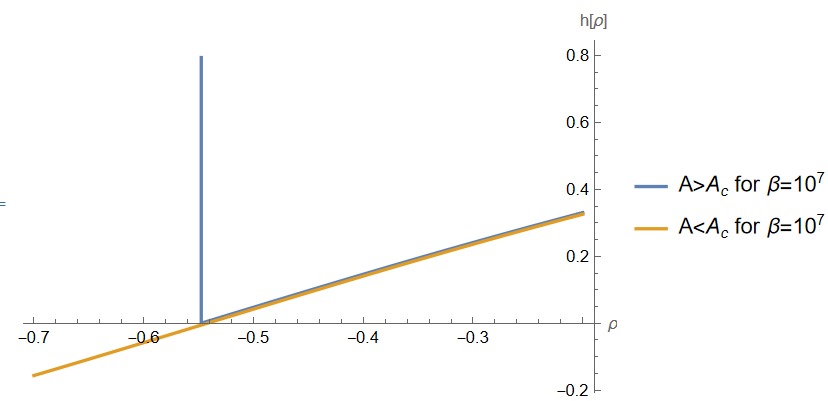}
         \caption{}
         \label{fig:hl60}
     \end{subfigure}
     \hfill
     \begin{subfigure}[b]{0.45\textwidth}
         \centering
         \includegraphics[width=\textwidth]{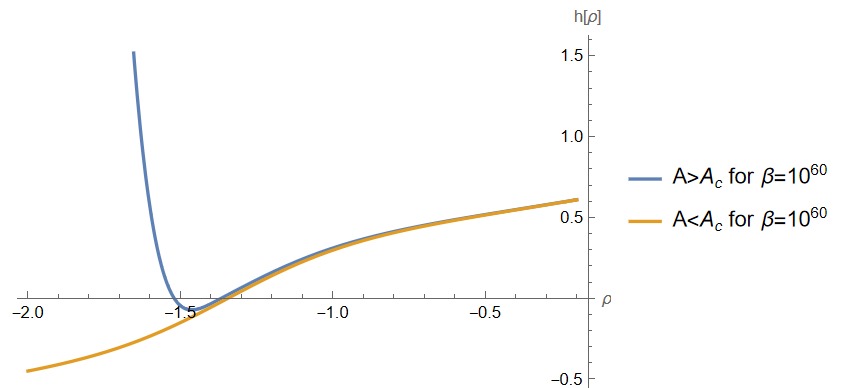}
         \caption{}
         \label{fig:hl7}
     \end{subfigure}
     \caption{With increase in $\beta$, the transition from $A<A_c$ to $A>A_c$ becomes less abrupt. Note that the dependence on $\beta$ is very weak, even though the trend that the trough of $h$ is getting pushed out to more negative $\rho$ should be clear. For doing numerics, $\log g$ and $\log \beta$ are better suited as variables.}
     \label{gig}
\end{figure}

The integral depends only on the boundary values of the integrand since it is a total derivative. Note that we expect the first order system to be most directly useful in the cap region, which corresponds to large $\beta$. In this regime, the outer (upper) boundary conditions are at $\rho \rightarrow 0$, and can be found from \eqref{hgexp}:
\begin{equation}
\label{upperlimit}
    h'(\rho \rightarrow 0) = \frac{2 \pi}{\beta}, 
    \qquad
    g (\rho \rightarrow 0)= \frac{\beta^2}{16 \pi^2}
\end{equation}
At the other end of the integral (i.e., $\rho \rightarrow -\infty$ limit), the series expansions for $h$, $\Phi$ and $\chi$ are the same as in the 1+1 dimensional case (see eqns. (3.4-3.6) of \cite{ItzhakiPuncture}). This can be directly checked using a systematic series expansion, but it is also easy to see -- in this limit, the $g(\rho) \rightarrow$ large constant, and therefore the system reduces to the 1+1 form. Seeing this numerically in the $h$-plots is harder, because the dependence on $\beta$ is quite slow -- we present some plots in Figure \ref{gig}. In any event, $h' \to 0$ as $\rho \to - \infty$. 

The final expression for the entropy of the winding condensate becomes
\begin{equation}
    S_W = \frac{4 \pi \beta}{\kappa_0^2} \times \frac{2 \pi}{\beta} \times  \frac{\beta^2}{16 \pi^2} e^{-2 \Phi_0} = \frac{\beta^2}{2 \kappa_0^2} e^{-2 \Phi_0}
\end{equation}
where $\Phi_0$ is the value to which the dilaton saturates at the cap. Newton's constant in Einstein's frame $G_N$ is related to $\kappa_0$ and the constant dilaton $\Phi_0$ as
\begin{equation}
    8 \pi G_N = \kappa_0^2 e^{2 \Phi_0},
\end{equation}
resulting in
\begin{equation}
    S_W = \frac{\beta^2}{16 \pi G_N}.
\end{equation}
This is precisely the Bekenstein-Hawking entropy of a 3+1 black hole with mass $m$ such that $\beta = 8 \pi G_N m$, but here we obtained it from the winding condensate entropy.

\section{A ``Decoupling" of the Cap Region?}\label{Conclusion}

Our modest goal in this paper was to see how much of the discussion in \cite{ItzhakiPuncture} generalizes to general dimensions. We found that essentially all of it does, {\em if we are working in the ``cap limit"}. There were two key technical ingredients that enabled this. The first, was that the Horowitz-Polchinski equations allowed a re-writing in terms of a first order system for the metric-dilaton-winding sector in general dimensions, despite the complications introduced by the sphere. Secondly, the resulting equations have a universal structure in the ``cap limit" which can be viewed as a large-mass approximation near the horizon. In this limit, the equations reduced in a precise sense to the first order HP system in 1+1 dimensions. These two facts were sufficient for us to generalize all of the successes of \cite{ItzhakiPuncture} to higher dimensions. These results suggest that the cap region physics of the 1+1 black hole, is a universal lesson from 1+1 dimensions to more general black holes.

As always, these results raise new questions and puzzles. We list some of them below and make some related comments and speculations.
\begin{itemize}
    \item It will be interesting to understand the precise connection between the entropy calculation here (also  \cite{ItzhakiPuncture}) and that in \cite{Brustein}. The latter calculation managed to reproduce the entropy, but it was done in the second-order HP system and did not rely on the critical solution. It is interesting to note that in our case, the argument for dropping the contribution from the lower end of the winding integral relied on the details of the boundary conditions at the puncture, while the entropy is a kind of conserved ``charge" visible at any radial cut (including at the horizon) in the work of \cite{Brustein}\footnote{We thank Yoav Zigdon for a clarifying comment on this point.}. It is tempting to think that this is an Euclidean manifestation of the ``smooth horizon without a smooth horizon" suggestion in \cite{Vaibhav1, Vaibhav2} for Lorentzian black holes. The observation there was that a UV-complete bulk description without a manifest interior can approximate (exponentially well!) the smooth horizon correlation functions, thanks to crucial properties of black hole normal modes\footnote{Key features of these modes that enable this, were first noted numerically (in a different context) in \cite{Preetham}. See various follow-up works in \cite{2301.11780, 2308.11704, 2312.14109, 2410.00732, 2310.06438, 2401.01417, 2411.09500, 2406.10879, SumanAdS5}.}. 

    \item It was noted in \cite{ItzhakiPuncture} that the puncture at the tip may be used to evade the Witten index obstruction noted in \cite{ChenMaldacenaWitten} when connecting black holes to strings. Our calculation in higher dimensions is consistent with this.
    
    \item The cap region (in a language adapted to 1+1 dimensions) is defined by 
    \bea
    \rho \ {\lesssim} \ \sqrt{k \alpha'}
    \eea
    after re-instating string length. In particular, this includes the region $\rho \sim \sqrt{\alpha'}$. The original BPS D-brane decoupling limit of Maldacena zoomed in on this latter region, and it was formulated in Lorentzian signature. The Euclidean cap region on the other hand is aware of the stack size of the NS5-branes (note that $k$ is the number of NS5-branes in this language). This and the finite temperature of the system are reminiscent of a deconfined version of the decoupling limit -- it will, of course, be interesting to make this speculation precise.

    \item It can be checked that the 3+1 first-order HP system does not have a solution that goes over to conventional Schwarzschild in the asymptotic region, with an asymptotically constant dilaton. This is again a suggestion that the significance of FZZ duality is in the cap region. It will be very interesting to understand what extra ingredient is provided by the orthogonal complement of the first order system, within the solution space of the second order system, that allows extrapolation to the asymptotic region. This complement seems to be an important ingredient in gluing the cap to the asymptotic region. We suspect that the first order system is a rough analogue of the normalizable mode in the context of AdS/CFT. A hint of this is provided by the observation made below \eqref{chibc} that the first order system retains the normalizable mode of the second order system. Understanding this better, may shed light on the construction of the Euclidean (and perhaps Lorentzian) Schwarzschild solutions in string theory. These ideas are likely to be of significance in understanding the first bullet point in this section as well. 
    
    \item It was observed in \cite{Soda, Emparan, ChenMaldacena} that the large-$D$ limit of the Schwarzschild black hole naturally leads to the 1+1 dimensional black hole. The discussion there (see eqn. (6) of \cite{ChenMaldacena}) is naturally in the $\beta \gg 1$ limit, which corresponds to our cap region.
    \item A natural expectation from the winding condensate picture around the shrinking Euclidean circle is that the wave function of the winding mode goes as 
    \bea
    \chi(\rho) \sim e^{-S_{NG}(\rho)}
    \eea
    where $S_{NG}$ is the Nambu-Goto action associated to a string worldsheet wrapping the tip of the cigar up to the radius $\rho$. It turns out that this is precisely reproduced by (say) our \eqref{4final2}. Upon integrating, \eqref{4final2} leads to $\chi(\rho) = A e^{-\frac{\beta}{2 \pi} \int_0^\rho h(\rho') d\rho'}$, where we have re-instated the $\tilde \beta$ that was scaled out from the 1st order system. This expression is reproduced by $S_{NG} = \frac{1}{2\pi}\int_0^{\beta} \int_0^\rho d^2\sigma\sqrt{\gamma}$ where $\gamma_{ab}=g_{\mu \nu}\frac{\partial{X^{\mu}}}{\partial \sigma^a}\frac{\partial{X^{\nu}}}{\partial \sigma^b}$, with $g_{\mu\nu}$ the spacetime metric \eqref{dmetric} and $\sigma^a=(\sigma^1, \sigma^2)$, the worldsheet coordinates. The natural worldsheet embedding in target space is $(X^0\equiv\tau=\sigma^1, X^1\equiv\rho=\sigma^2$) and the rest of $X^i={\rm const.}$, so that $\gamma_{ab}={\rm diag}(h^2(\rho),1)$. Here $X^i$ are the coordinates on the spacetime sphere. 
\end{itemize}

\section*{Acknowledgments}

 CK thanks Pallab Basu and Yoav Zigdon for discussions \& correspondence and the University of Witwatersrand for hospitality during the early stages of this work. ST acknowledges the support of the KVPY scholarship (DST, Govt. of India), which, quite frustratingly is discontinued, and also thanks Sabyasachi Pramanik for providing accommodation at the IISc campus even after his graduation during the later stages of this work.   

\appendix

\section{Derivation of First Order System in $D$ dimensions}
\label{find1st}

We repeat the second-order system here for convenience:
\begin{equation}\label{deom1}
\begin{split}
    2(\Phi''-\Phi'^2) + \frac{(D-2)(D-3)}{8}\l(\frac{g'}{g}\r)^2 &- \frac{(D-2)(D-3)}{2g}+
    \\
    &- \Lambda+\l(\frac{D-2}{2}\r)\l(\frac{g'h'}{gh}\r) = \chi'^2+\chi^2(3\Tilde{\beta}^2 h^2-\Tilde{\beta_H}^2),
\end{split}   
\end{equation}
\begin{equation}\label{deom2}
    \frac{D-3}{g}- \frac{1}{2}\frac{g'h'}{gh}-\frac{g''}{2g}+\frac{g'\Phi'}{g} - \frac{D-4}{4} \l(\frac{g'}{g}\r)^2=0,
\end{equation}
\begin{equation}\label{deom3}
     \frac{e^{2\Phi}}{g^{\frac{D-2}{2}}h}\partial_{\rho}(g^{\frac{D-2}{2}}h\chi'e^{-2\Phi})=\left(\Tilde{\beta}^2 h^2-\Tilde{\beta_H}^2\right)\chi,
\end{equation}
\begin{equation}\label{deom4}
\begin{split}
    h\left(\frac{\Phi'}{h}\right)'+ \frac{(D-2)(D-3)}{8}\l(\frac{g'}{g}\r)^2 &+ \l(\frac{D-2}{2}\r)\l(\frac{g'h'}{gh}\r) + \\
    &- \frac{(D-2)(D-3)}{2g} - \frac{D-2}{2} \frac{g' \Phi'}{g}=\chi'^2+\Tilde{\beta}^2 h^2\chi^2
\end{split}
\end{equation}
We will look for a subsystem where 
\begin{equation}
\label{first1}
    \chi' = - \tb h \chi.
\end{equation}
Plugging this in \eqref{deom3} immediately gives
\begin{equation}
\label{first2}
    h' = h \l(\Phi'- \frac{D-2}{4} \frac{g'}{g}\r) + \frac{\tb_H^2}{2 \tb}.
\end{equation}
Use the following ansatz (we will determine $A$ in what follows):
\begin{equation}
\label{ansatz}
    \frac{\Phi'}{h} = - \tb \chi^2 + A.
\end{equation}
Take derivative of \eqref{ansatz}, use \eqref{first2} to get
\begin{equation}
    2 (\Phi'' - \Phi'^2) + \frac{D-2}{2} \frac{g' \Phi'}{g} - \frac{\tb_H^2}{\tb} \frac{\Phi'}{h} = 4 \tb^2 h^2 \chi^2 + 2 A' h
\end{equation}
Subtracting it from \eqref{deom1} we find
\begin{equation}
\label{inter}
    \frac{D-2}{2} \frac{g' \Phi'}{g} - \frac{\tb_H^2}{\tb}A - \frac{(D-2)(D-3)}{8}\l(\frac{g'}{g}\r)^2 + \frac{(D-2)(D-3)}{2g}+ \Lambda - \frac{D-2}{2} \l(\frac{g'h'}{gh}\r) = 2 A' h
\end{equation}
To get an expression for $A'h$, take derivative of \eqref{ansatz}, use \eqref{first1}, and subtract from \eqref{deom4}. The result is: 
\begin{equation}
    A'h = \frac{D-2}{2} \l(\frac{g' \Phi'}{g} + \frac{D-3}{g} - \frac{g'h'}{gh} - \frac{D-3}{4} \l(\frac{g'}{g}\r)^2\r).
\end{equation}
Plugging this in \eqref{inter} yields
\begin{equation}
\label{A}
    A = \frac{\tb}{\tb_H^2} \l(\frac{D-2}{2} \frac{g'h'}{gh} + \frac{(D-2)(D-3)}{8} \l(\frac{g'}{g}\r)^2 - \frac{D-2}{2} \frac{g' \Phi'}{g} - \frac{(D-2)(D-3)}{2g} + \Lambda\r)
\end{equation}
which finally gives, after using \eqref{first2} once
\begin{equation}
    \label{first3}
    \frac{\Phi'}{h} = -\tb \chi^2 + \frac{\tb}{\tb_H^2} \l(\frac{D-2}{4} \frac{\tb_H^2}{\tb} \frac{g'}{gh} - \frac{D-2}{8} \l(\frac{g'}{g}\r)^2 - \frac{(D-2)(D-3)}{2g} + \Lambda\r).
\end{equation}
Together \eqref{first1}, \eqref{first2}, and \eqref{first3} are the first order system for the winding mode, metric component $h$ and the dilaton. The calculations here were more involved than in \cite{ItzhakiPuncture} due to the presence of $g$, but gratifyingly, the key structures are preserved and all the complications could be separated into the metric function $g$.

\bibliographystyle{unsrt}
\bibliography{bibliography}

\end{document}